\documentclass[aip,reprint]{revtex4-1}
\usepackage{graphicx}
\usepackage{dcolumn}
\usepackage{bm}
\usepackage{mathtools}
\draft 
\begin{document}
\title{Energy-transfer Quantum Dynamics of HeH$^+$  with He atoms: Rotationally Inelastic Cross Sections and Rate Coefficients}

\author{F. A. Gianturco}
\affiliation{Institut f\"{u}r Ionenphysik und Angewandte Physik, Universit\"{a}t Innsbruck\\
Technikerstr. 25
A-6020, Innsbruck, Austria}

\author{K. Giri}
\affiliation{Department of Computational Sciences, Central University of Punjab,\\
Bathinda 151001 India}

\author{L. Gonz\'{a}lez-S\'{a}nchez}
\email{lgonsan@usal.es}
\affiliation{Departamento de Qu\'{i}mica F\'{i}sica, University of Salamanca\\
 Plaza de los Ca\'{i}dos sn,
  37008, Salamanca, Spain}

\author{E. Yurtsever}
\affiliation{Department of Chemistry, Koç University \\
Rumelifeneriyolu, Sariyer
TR 34450, Istanbul, Turkey}

\author{N. Sathyamurthy}
\affiliation{Indian Institute of Science Education and Research Mohali\\
SAS Nagar, Manauli 140306 India}

\author{R. Wester}
\affiliation{Institut f\"{u}r Ionenphysik und Angewandte Physik, Universit\"{a}t Innsbruck\\
Technikerstr. 25
A-6020, Innsbruck, Austria}

\date{\today}

\begin{abstract}
\hspace*{6cm}{\textbf {Abstract}}

Two different $ab$  $initio$ potential energy surfaces are employed to
investigate  the efficiency of the rotational excitation channels  for the polar molecular ion HeH$^+$ interacting with He atoms. We further use them to  investigate the quantum  dynamics of both  the proton-exchange reaction and the purely rotational
inelastic collisions over a broad range of temperatures. In current modeling studies, this cation is
considered to be one of the possible cooling sources under early universe  conditions after the recombination era and has recently been found to exist  in the Interstellar Medium.  Results from the present calculations are able to show
the large efficiency of the  state-changing channels involving   rotational states of this cation. In fact, we find them
to be similar in size and behaviour to the inelastic and to the reaction rate coefficients obtained in 
previous studies where H atoms  were employed as projectiles. The same rotational excitation 
processes, occurring when free electrons are  the collision partners of this 
cation, are also compared  with the present findings. The relative importance of the reactive, proton-exchange channel and the purely inelastic channels is also  analysed and discussed. The  rotational de-excitation processes are also investigated for the cooling kinetics of the present cation under cold trap conditions with He as the buffer gas.  The implications of the present results for  setting up more comprehensive 
numerical models to describe the chemical evolution networks in different environments are briefly discussed.

\end{abstract}


\maketitle


\begin{large}
{\Large}
\begin{large}

\end{large}\end{large}
\section{Introduction} \label{sec: intro}
   Since the detection of HeH$^+$ in the planetary nebula NGC 7026 by
\citet{Gusten2019} and its confirmation by 
\citet{Neufeld2020}, there has been a renewed interest in the
mechanism of formation and destruction of HeH$^+$ in stellar and interstellar conditions. 
\citet{Novotny2019} have recently measured the recombination rate for HeH$^+$ in the ion storage ring and found such rates, which lead to the destruction of HeH$^+$, to be much smaller than what was expected 
from earlier studies, thus suggesting that this important cation should be more abundant than previously 
expected in the astrochemical environments: from circumstellar envelopes to the stage of the recombination era in the early universe. Hence, we have witnessed a revival of the interest  in discussing and modeling its collision 
efficiency in operating as an energy dissipation partner with other  chemical species like He, H, 
H$^+$ and H$_2$, all partners considered by several modeling studies to be present in similar environments.\citep{Galli_Palla2013, LeStDa02}
\\

The molecular cation HeH$^+$ was detected in the laboratory using a mass spectrograph
as early as 1925. \citep{Hogness_Lunn1925} The infrared spectrum 
 was predicted  by Dabrowski and Herzberg in 1977. \citep{Dabrowski_Herzberg1977} 
But HeH$^+$ eluded detection in interstellar conditions until rather recently, 
as mentioned earlier. \citep{Gusten2019}\\

Our understanding so far is that soon after the nucleosynthesis was over and
conditions were conducive to recombination processes 
(red shift z $\sim$1000 and temperature $\sim$3000-4000 K),
helium, hydrogen and to a less extent lithium atoms were formed. \citep{Galli_Palla2013, LeStDa02} 
Although earlier studies \citep{Bates1951} had proposed the formation of H$_2^+$ by 
radiative association (RA) from H and H$^+$, subsequent studies suggested the
formation of HeH$^+$ as the first molecular species by the following route:
\begin{equation}
\textrm{He} + \textrm{H}^+ \longrightarrow \textrm{HeH}^+ + \textrm{h}\nu 
\end{equation}
However, the formation of HeH$^+$ under interstellar conditions is attributed to another
RA channel. \citep{Forrey2020}  That is,
\begin{equation}
\textnormal{He}^+ + \textnormal{H} \longrightarrow \textnormal{HeH}^+ +  \textrm{h}\nu 
\end{equation}
HeH$^+$ could react readily with the abundant H atoms (He : H = 1 : 10 in the early universe)
and form H$_2^+$:
\begin{equation}
\textnormal{HeH}^+ + \textnormal{H} \longrightarrow \textnormal{He} + \textnormal{H}_2^+. 
\end{equation}
The interaction of HeH$^+$ with H$_2$ to give HeH$_3^+$ and also possibly
He and H$_3^+$  have also  been considered by Zicler et al. \citep{Zicler2017} but the corresponding  inelastic collisions leading to rotational or rovibrational energy transfer have not been included explicitly within that chemical network.

Despite the difficulties in detecting HeH$^+$ in the planetary nebula, \citet{Gusten2019} pointed out that the
recorded emission of HeH$^+$ from the rotational state $j$ = 1 to $j$ = 0  is (2-3 times) larger than 
what could be accounted for by the available rate coefficients and astrophysical models. 
\citet{Ravi2020} have proposed recently that the nonadiabatic coupling terms between
He, H and H$^+$ could act like (astronomical) friction and form [HeH$_2^+$]*, which could in turn  lead to
the formation of HeH$^+$ and H and also of He + H$_2^+$. All the above considerations point out to the  relevance that the present molecular cation is currently expected to have within the chemistry of the early universe  and of the interstellar medium (ISM), as we shall further illustrate below.    \\

Inelastic collisions between HeH$^+$ and H and the reactive events between the two have also been investigated
extensively over the years \citep{Bovino2011, Fazio2014, Lique2020}, and the reverse reaction 
\begin{equation}
\textnormal{He} + \textnormal{H}_2^+ \longrightarrow \textnormal{HeH}^+ + \textnormal{H} 
\end{equation}
has been studied extensively over the last several decades (for example, see \citet{Ramachandran2009, Kolakkandy2012, Fazio2014}). Some of their results will be also discussed in the following.\\
\

      More specifically,  in the present work we intend to investigate in some detail, and to our knowledge for
the first time, how efficiently HeH$^+$ could  also exchange internal energy (mainly rotational energy) when it  interacts with the neutral He atoms   known to be  present in the
same environments, and therefore enter the paths in  energy dissipation networks  by undergoing purely inelastic (rotational)  and H$^+$-exchange processes with that neutral partner:
\begin{equation}
\label{eq:prrxn}
\textnormal{HeH}^+{(\nu, j)} + \textnormal{He}^{'} \longrightarrow \textnormal{HeH}^+{(\nu, j')} + \textnormal{He}^{'}.
\end{equation}
\begin{equation}
\label{eq:prrin}
\textnormal{HeH}^+ + \textnormal{He}^{'} \longrightarrow \textnormal{He} + \textnormal{H}^+\textnormal{He}^{'}.
\end{equation}

\
We have labelled one of the two helium atoms as He$^{\prime}$ in equations (\ref{eq:prrxn}) and (\ref{eq:prrin}) to distinguish between the two atoms when we shall also be considering  the H$^+$- exchange collision events, as further discussed below. 

       The equilibrium geometry and the potential well depth for  He$_2$H$^+$ has been investigated over the years by several
workers. \citep{Poshusta1969, Poshusta1971, Milleur1974, Dykstra1983, Baccarelli1997, Filippone_Gianturco1998, kim_Lee1999}  
Although a limited number of geometries near the equilibrium geometry of He$_2$H$^+$ was investigated by \citet{Dykstra1983} and
an analytical fit of the potential energy surface (PES) was given by \citet{Lee1986} the first extensive $ab$ $initio$ PES
for the system was generated by \citet{Panda_Sathyamurthy2003} using the coupled cluster singles
and doubles excitation with perturbative triples (CCSDT(T)) method employing the d-aug-cc-pVTZ basis set. A deep potential
well of depth 0.578 eV was reported for the collinear geometry [He-H-He]$^{+}$ with a He-H distance of 0.926 \AA\ and several bound and quasi-bound states
(for total angular momentum $J$ = 0) were determined in that work. In the present report, their potential function is referred to as PS-PES. 

Using the time-dependent quantum mechanical
wave packet (TDQMWP) method \citep{Balakrishnan1997}, \citet{Bhattacharya_Panda2009} investigated the H$^+$-exchange reaction
in (HeH$^+$/ He) collisions for different vibrational ($v$) and rotational ($j$) states of HeH$^+$ and found that there
were severe oscillations in the plots of the exchange reaction probability as a function of the relative kinetic
energy ($E_{\textrm{trans}}$) of the reactants. These oscillations could be traced to the bound and quasi-bound vibrational
states of HeH$^+$...He complex. \citep{Panda_Sathyamurthy2003} Although the oscillations were considerably quenched in the plots of partial reaction
cross section as a function of $E_{\textrm{trans}}$, some of the oscillations survived in the plots of the corresponding  excitation function. \citep{Panda_Sathyamurthy2003}
Additionally, \citet{Liang2012} reported another $ab$ $initio$ PES for the system, hereinafter referred to as the LYWZ PES, obtained using the multi-reference configuration
method and the d-aug-cc-pV5Z basis set. It was comparable to the PS-PES in all its general and specific features, as noted in  \citep{Liang2012}, so we shall be using the PS-PES in the present study. 
  These authors  \citep{Liang2012} also carried out quasi-classical trajectory calculations using their LYWZ PES and
found that the resulting excitation functions were also comparable in magnitude to those reported by \citet{Panda_Sathyamurthy2003},
except for small oscillations in the latter which were not present in \citep{Liang2012}. Furthermore, by using the TDQMWP approach, \citet{Xu_Zhang2013} additionally computed the integral reaction
cross sections on the LYWZ PES over a range of $E_{\textrm{trans}}$ (0-0.5 eV) and pointed out the importance of including
Coriolis coupling. The work of  \citet{Wu2014} extended the above  study and computed state-to-state differential and integral reaction
cross sections on the same PES. They also reported rate coefficients for the exchange reaction for $v$ = 0, $j$ = 0 of
HeH$^+$ ranging from 7x10$^{-11}$ to 3x10$^{-10}$ cm$^3$ molecule$^{-1}$ s$^{-1}$ over a temperature range
of 0-200 K. Yao \citep{Yao2014}  studied the dynamics of the HeD$^+$ + He exchange reaction. Unfortunately, none of these studies
investigated the anisotropy of the potential around HeH$^+$ and purely inelastic vib-rotational processes in (HeH$^+$/ He) collisions.

Liang et al. \citep{Liang2012} had shown that the two PESs are nearly identical and  therefore, we shall restrict our present analysis by carrying out rotational inelastic calculations  using the PS-PES. \citep{Panda_Sathyamurthy2003}, together with the calculations which make use of the newly computed RR-PES discussed in the following Section.

The best available theoretical value for the dipole moment  is $\mu$ = 1.66 D as given by  \citet{05PaBuAd} while the  rotational constant is B = 33.526 cm$^{-1}$ as quoted in \citet{05MuStWi}. These are the values employed in the present work.

We have also carried out a different set of  $ab$ $initio$ calculations by generating a new PES which is focused on the purely inelastic collisions without considering the H$^+$-exchange channel mentioned above. The target molecule was therefore taken to be at a fixed internuclear distance given by its equilibrium value (see below) and will be called a rigid-rotor (RR) potential energy surface.  The purely inelastic cross sections generated by using this new PES will also be compared with those obtained from the dynamics where the H$^+$-exchange channels were also included when using the earlier PS-PES. The following section will present in more details the features of both $ab$ $initio$  calculations, while later on their dynamics will be discussed and compared.

One of the reasons for the present study is linked to the notion that is  of direct interest to have quantitative information on the relative efficiency of a variety of
 energy-changing processes involving the internal level structure of the HeH$^+$ polar cation when it is made to interact with other "chemical" partners  considered to be present in a variety of interstellar environments. Thus, it is important to know how possible partners   like H and He neutral atoms, or the  free electrons, are affecting internal  energy redistribution in the molecular cation to make it a significant partner for the general cooling channels deemed important  following the recombination era \citep{Galli_Palla2013} . Hence, the above question is central to the conclusions of our present investigation reported in the last Section.
 
  Our  present results will therefore   be compared with those already available for (HeH$^+$/ H) \citep{Lique2020} and  for
(HeH$^+$/ e$^-$) \citep{Hamilton2016, Ayouz2019} collisions leading to rotational excitations of the molecular cation.  As we shall show below, one of the  important findings of our present study is that the neutral helium
atoms turn out to be as efficient as, if not more than, hydrogen atoms in causing rotational excitation in HeH$^+$ and therefore their corresponding inelastic rate coefficients should be included in the kinetics modeling the chemical evolution in early universe and ISM environments.
\\

The newly constructed $ab$ $initio$ PES for the rigid rotor HeH$^+$-He interaction is described in section \ref{sec:pes} and compared there
with the available PS-PES. The methodology adopted for the investigation of the inelastic processes is described in 
sections \ref{sec:rrd} and \ref{sec:psd}, while the computed results of inelastic cross sections and rate coefficients are
presented, discussed and compared with  available results involving other collision partners  
in section \ref{sec:resultd}. Einstein spontaneous emission coefficients are also presented and discussed in relation to   generating critical density values under different  early universe conditions to estimate the relative importance of the collisional and radiative decays, and the results are reported in  section \ref{sec:erc}. A brief description of the quantum rotational kinetics of HeH$^+$ in a cold ion trap with He as a buffer gas will be given in  \ref{sec:rrk}. A summary of our findings and their importance for chemical network 
modelings  of the cooling role of the title cation will finally be  presented in section \ref{sec:pconc}.

\section{Ab initio interaction potentials} \label{sec:pes}


For the HeH$^+$/He system, new and extensive \textit{ab initio} calculations were carried out in the present work using
the MOLPRO suite of quantum chemistry codes: see \citet{MOLPRO,MOLPRO_brief}. 
\begin{figure}[!ht]
        \begin{center}
                \begin{tabular}{cc}
                        \vspace*{-14mm} \resizebox{9.7cm}{!}{\includegraphics[angle=270]{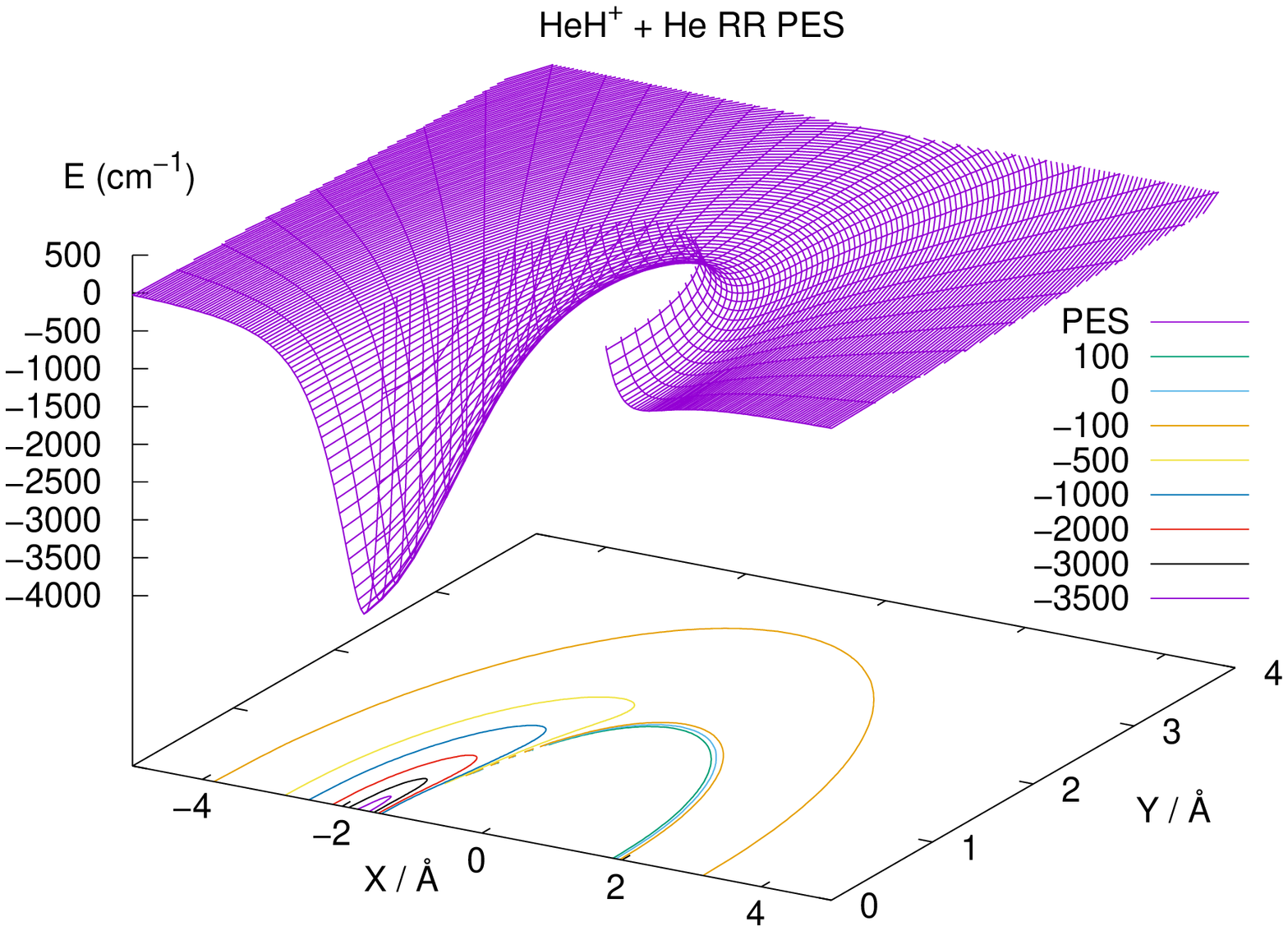}}\\
                        \resizebox{9.2cm}{!}{\includegraphics{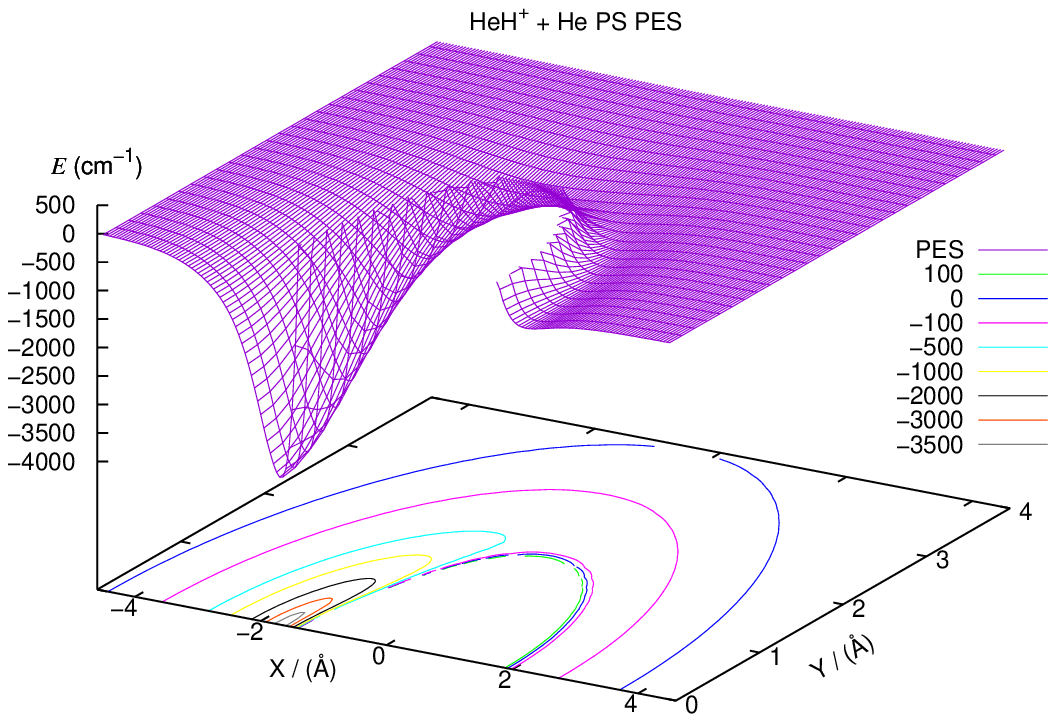}} \\
                \end{tabular}
                \caption{3D perspective plots of the two PESs for rigid rotor HeH$^+$ ($r$ = 0.774 \AA) interacting with He, and  potential energy contours in ($R$, $\theta$) space. top panel: results from the RR-PES computed in this work;  bottom panel: the same rigid-rotor PES   obtained from the  reactive PS-PES discussed both earlier and in this Section.}
                \label{fig:fig1}
        \end{center}
\end{figure}
The HeH$^+$ bond distance was kept fixed at
0.774 \AA\ throughout the calculation of the potential energy points, thereby producing the RR-PES. 
The post-Hartree-Fock treatment was carried out using
the CCSD(T) method as implemented in MOLPRO \citep{92HaPeWe,94DeKnxx} 
and complete basis set (CBS) extrapolation using the aug-cc-pVTZ, aug-cc-pVQZ and aug-cc-pV5Z basis sets
\citep{93WoDuxx,94WoDuxx} was carried out.
The basis-set-superposition-error (BSSE) \citep{70BoBexx} was also included for all the calculated points so that 
the full interaction was obtained with the inclusion of the BSSE correction.

\begin{figure}[!ht]
        \includegraphics[width=0.49\textwidth]{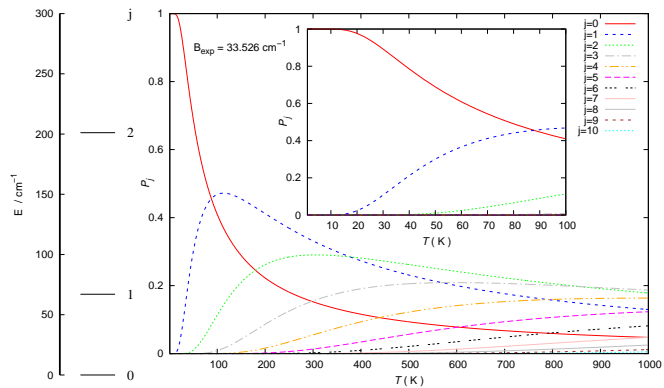}
        \caption{Energy spacings between the lower rotational levels of the present cation (left) and steady-state distribution of the
	relative populations among rotational levels as a function of temperature (right in figure).}
        \label{fig:fig2}
\end{figure}

The two dimensional (2D) RR-PES ($R$, $\theta$) was
calculated using 76 points from 1.0 to 10.0 \AA\  along $R$ and 19 values from 0 to 180$^\circ$ in  $\theta$ for a total
of 1,444 grid points. We report in Figure \ref{fig:fig1} a pictorial representation of the new RR-PES, 
given in 3D space and also projected below it as potential energy contours, in the upper panel of the figure.  The lower panel reports the results for the same rigid-rotor two-dimensional reduction of the more complete 3D reactive potential given by the PS-PES already discussed in the previous Section.  It is interesting to note that both interactions exhibit a deep attractive well on the linear geometry forming the (He-H-He)$^+$ complex, as already discussed in much of the earlier work mentioned in the Introduction and analysed  in the more recent studies on the stable HeHHe$^+$ molecule \citep{FW20, SF17}.

This pictorial comparison of the two PESs that will be employed to generate rotationally inelastic cross sections and rate coefficients in the next Section clearly reveals that they are largely identical, a feature which will be further  discussed  below. The He-end of the cation is located on the positive side of the X coordinate in the figure panels.

It is also useful at this point  to  provide a pictorial view of the structural properties of the HeH$^+$ molecular
ion in terms of the energy spacing between its lower rotational levels and the resulting steady-state distributions
of their relative populations over a range of temperatures relevant for the present discussion. They are reported
in Figure \ref{fig:fig2} where one clearly sees how only relatively few states would be populated under equilibrium conditions up to 1000 K. These features will  play a role when discussing
the inelastic rate coefficients  from the quantum dynamics in the following Section.

The original data points of the new RR-PES calculations were used by interpolation
to  generate a grid with 601 radial points (between 1 and 10 \AA) and 37 angular values (from 0 to 180$^\circ$).
\begin{figure}[!ht]
        \includegraphics[width=0.365\textwidth,angle=270]{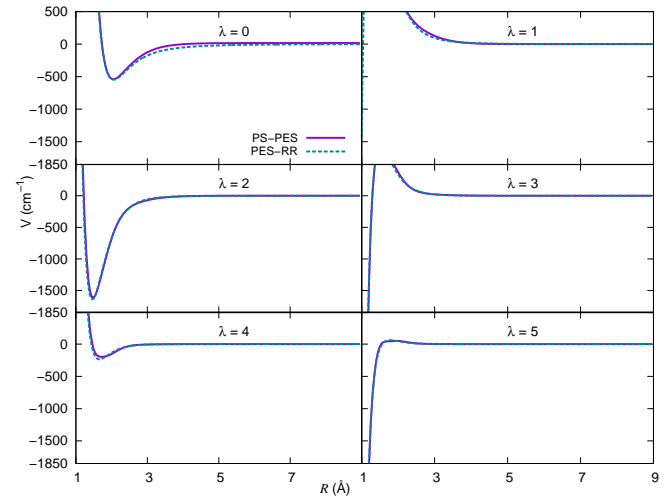}
        \caption{Multipolar expansion coefficients computed for the RR-PES interaction of
                the present study and from the 2D-reduced representation of the earlier PS-PES also employed in the present study. See main text for further details.}
        \label{fig:fig3}
\end{figure}
\\
For a  more direct, and quantitative evaluation  of  the spatial anisotropy of this PES, it is also useful to  expand this extensive 2D grid 
for the fixed-bond (RR) target geometry in terms of Legendre polynomials as given below:

\begin{equation}
\textnormal{V}(R, \theta) = \Sigma_{\lambda} V_{\lambda}(R)P_{\lambda}(cos\theta)
\end{equation}
\
We have  initially obtained 25 multipolar coefficients for each of the involved rigid-rotor PESs, although only the first 10 were actually 
included in the  scattering calculations discussed later 
since they turned out to be sufficient to reach numerical convergence of the cross section values.The calculated radial coefficients exhibited a  root-mean-square error to the initial RR-PES points of about 0.64 cm$^{-1}$ along the radial range. The obtained radial coefficients were interpolated during their usage in our in-house scattering code (see below) and further extrapolated using the asymptotic representations of the lower  coefficients, thus  ensuring that the overall interaction includes 
asymptotically the leading dipole polarizability of the neutral atomic partner plus the dipole-polarization terms:

\begin{equation}
	V_{lr}(R,\theta) \sim \frac{\alpha_0}{2R^4} + \frac{2 \alpha_0 \mu}{R^5} \cos{\theta}
\label{eq1}
\end{equation}

where $\alpha_0$=1.41 a$^3_0$ is the polarizability of the helium atom and $\mu$ is the permanent dipole of the HeH$^+$ partner reported earlier.

The resulting radial coefficients from the above expansion, for both the new RR-PES and for the 2D-reduction of the fuller 3D PES that
includes the reactive channels and which we shall analyse in the next subSection, the PS-PES,
are compared in the panels  of Figure  \ref{fig:fig3}, where only the first six of them are actually shown.  

Here we must mention specifically that Panda and Sathyamurthy\citep{Panda_Sathyamurthy2003} had reported a global analytical
potential energy function for the PS-PES. Therefore, keeping the equilibrium bond distance ($r_{e}$) of HeH$^+$  fixed
at 0.774 \AA, we compute $V_{\lambda}(R)$ numerically by integrating  $\textnormal{V}(R, r_{e}, \theta)$ over $\theta$:
\begin{equation}
\textnormal{V$_\lambda$}(R) = \frac{(2\lambda + 1)}{2} \int_{-1}^{1} V(R, r_e, \theta)P_{\lambda}
(cos\theta)d(cos\theta)
\end{equation}

 One clearly sees from that
comparison that, in terms of the 2D description which will be employed to calculate the purely rotationally inelastic dynamics
in the present work, the two potential functions behave very closely and are therefore  likely to yield very similar inelastic
cross sections and corresponding inelastic rate coefficients as will be discussed in detail in the following Sections.

\section{ Rotational Inelastic dynamics on the RR-PES} \label{sec:rrd}

 We briefly report below the computational method employed in this work to obtain  purely rotationally inelastic cross sections and rate coefficients for the 
scattering of HeH$^+$ with He using the the two dimensional  RR-PES discussed in the preceding subSection.
The standard time-independent formulation of the Coupled-Channel (CC) approach to quantum  scattering has been known for many years already
(see for example \citet{Taylor2006} for a general text-book formulation) while the more recent literature on the actual computational methods has been also very large. For a selected set of references over the more recent years see for instance refs. \onlinecite{60ArDaxx, 79Secrxx, 97KoHoffxx, 94JmHxx, 79FaGxx}. However, since we have already discussed our specific computational methodology in many of our earlier 
publications \citep{03MaBoGi, 08LoBoGi, 15GoGiCa}, only a short outline of our approach will be given in the present discussion.

For the case where no chemical modifications are induced in the molecule by the impinging projectile, the total scattering wave function can be
expanded in terms of asymptotic target rotational eigenfunctions (within the rigid rotor approximation) which are taken to be spherical
harmonics and whose eigenvalues are given by $Bj(j + 1)$, where $B$ is the rotational constant for the closed-shell HeH$^+$ ion mentioned already in the previous Section:
33.526 cm$^{-1}$  and $j$ is the rotational quantum number. The channel components for the CC equations are therefore expanded into products of total angular momentum
$J$ eigenfunctions and of radial functions to be determined via the solutions of the CC equations \citep{03MaBoGi, 08LoBoGi} i.e. the 
familiar set of coupled, second order homogeneous differential equations:
\begin{equation}
\left(\frac{d^2}{dR^2} + \mathbf{K}^2 - \mathbf{V} - \frac{\mathbf{l}^2}{R^2} \right) \mathbf{\psi}^J = 0.
\label{eq:CC}
\end{equation}

In the above Coupled Equations, the $\mathbf{K}^2$ matrix contains the wavevector values for all the coupled channels of the problem and the $ \mathbf{V}$ matrix contains the full matrix of the anisotropic coupling potential.  The required scattering observables are obtained in the asymptotic region where the Log-Derivative matrix has a known form in terms of free-particle
solutions and unknown mixing coefficients. Therefore, at the end of the propagation one can use the Log-Derivative matrix to obtain the 
K-matrix by solving the following linear system:
\begin{equation}
(\mathbf{N}' - \mathbf{Y}\mathbf{N}) = \mathbf{J}' - \mathbf{Y}\mathbf{J}
\end{equation}
where the prime signs indicate radial derivatives, $\mathbf{J}(R)$ and $\mathbf{N}(R)$ are matrices of Riccati-Bessel and Riccati-Neumann functions. \citep{08LoBoGi} The matrix  $ \mathbf{Y}(R)$ collects the eigensolutions along the radial region of interest, out of which the Log derivative matrix is then constructed. \citep{08LoBoGi}
From the K-matrix produced by solving the coupled radial eq.s the S-matrix is then easily obtained and from it the state-to-state cross sections. \citep{08LoBoGi} We have already published an algorithm that
modifies the variable phase approach to solve that problem, specifically addressing the latter point and we defer the interested reader to
that reference for further details. \citep{03MaBoGi,08LoBoGi}

In the present calculations we have generated a broad range of state-to-state rotationally inelastic cross sections. 
The number of rotational states coupled within the dynamics was up to $j$ = 15 and the expansion over the $J$ values to converge the individual cross sections went up to $J$ = 100 at the highest energies. The radial range of integration during the propagation of the coupled eq.s
covered radial values from 1.0 to 1000.0 \AA\ using a variable number of points which went up to 5000 .
The range of $E_{\textrm{trans}}$  went from 10$^{-4}$ cm$^{-1}$ to 10$^{4}$ cm$^{-1}$  with 1500-2000 points for each considered transition.

Once the state-to-state inelastic integral cross sections ($\sigma_{j \rightarrow j'}$) are known, the rotationally inelastic rate coefficients $k_{j \rightarrow j'} (T)$  
can be evaluated as the convolution of the cross sections over a Boltzmann distribution of the  $E_{\textrm{trans}}$ values:

\begin{equation}
\begin{multlined}
k_{j \rightarrow j'} (T) =
\left(  \frac{8}{\pi \mu k_{\text{B}}^3 T^3}   \right)^{1/2} \\
\int_0^{\infty} E_{\textrm{trans}} \sigma_{j \rightarrow j'}(E_{\textrm{trans}}) 
 e^{-E_{\textrm{trans}}/k_{\text{B}} T} dE_{\textrm{trans}} 
\label{eq:rate}
\end{multlined}
\end{equation}

 The reduced mass value  
for the HeH$^+$/He system was taken to be 2.224975 a.u. The individual rate coefficients were obtained at intervals of
1K, starting from 5K and going up to 500K. The interplay between the changes in the reduced mass values, appearing in
the denominator in the equation above, and the structural strength within their corresponding PES will be further discussed
in the later Sections when the dynamical outcomes will be analysed.

\section{ Quantum Dynamics using both  the RR-PES and the PS-PES}\label{sec:psd}

Since the full 3D PS-PES discussed above allows for non-rigid rotor interaction as well as the H$^+$-exchange reaction, it is interesting at this point to evaluate the relative flux distributions going into the reactive and purely inelastic channels, so that a comparison can be made with the results produced by using the new RR-PES interaction. Hence, the (HeH$^+$/He) dynamics
was first  investigated using the ABC code \citep{Skouteris2000} for $v$ = 0, $j$ = 0 of HeH$^+$ over a range of $E_{\textrm{trans}}$ and including the presence of the H$^+$-exchange reactive channels.  It is important to mention the relevant code parameters used in our investigations, which cover a range of total energy (0.197 - 0.746 eV), with the maximum energy (Emax) = 2.15 eV in any channel and a maximum (Jmax) of 11 rotational states.  We have considered the maximum of hyperradius (Rmax) to be 24 $a_{0}$, with the number of log derivative propagation sectors (MTR) = 250, the total angular momentum (Jtotmax) going up to 100 and the helicity truncation parameter (Kmax) = 1.

From the same set of calculations we have  extracted the flux components going into the purely rotationally inelastic channels and we will discuss them  in comparison with the results for the same processes obtained using the RR-PES described in the previous Section. 

 The purely inelastic transition probability, as well
as the H$^+$-exchange reaction probability, have been computed for the $j$ = 0 $\rightarrow j^{'}$ = 1 process as an example of their behaviour. They  turned out, as expected, to markedly change   as a function of  $E_{\textrm{trans}}$ and
also of the contributing $J$. The tests of their convergence behaviour has been reported in  Figure S1 in the Supplementary Material, where it is also shown that  the oscillations which we have mentioned in the previous Section become markedly quenched when the partial cross sections and integral cross sections are plotted as a function of $E_{\textrm{trans}}$. 

\begin{figure}[!ht]
        \includegraphics[width=0.49\textwidth]{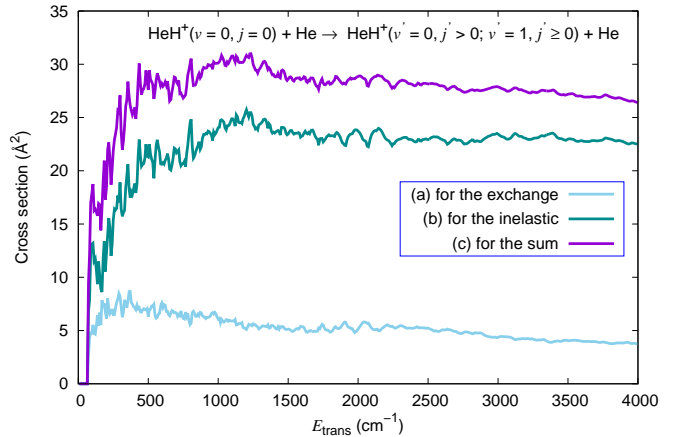}
        \caption{Integral inelastic cross section for (a) the H$^+$-exchange reaction leading to ${j^{\prime} > }$ 0, for ${v^{\prime}}$ = 0 and ${j^{\prime} \ge}$ 0, for ${v^{\prime}} > $  0, (b) the purely rotationally  inelastic process (${j^{\prime} > }$ 0) and (c) the sum of the two processes as a function of $E_{\textrm{trans}}$ for HeH$^+$ ($v$ = 0, $j$ = 0) + He collisions.}
        \label{fig:fig4}
\end{figure}

We now report in  Figure \ref{fig:fig4} the computed cross sections for the H$^+$- exchange reactions with $j^{\prime}$ $>$ 0:  they rapidly rise  starting from the same
threshold (67 cm$^{-1}$) at which  the inelastic channels open, while leveling off quickly with the  increasing of the relative $E_{\textrm{trans}}$.
The purely rotationally  inelastic cross sections reported in Figure \ref{fig:fig4} turn out to be
 significantly larger in magnitude than those obtained for the H$^+$-exchange channel. This may come as a surprise initially as the dynamics of the (HeH$^+$, He) system has a deep potential well of 0.578 eV and the dynamical outcomes are expected to be statistical (probability of exchange reaction $\sim$ probability of inelastic events $\sim$ 0.5).  However, it is known that the mere presence of a deep potential well does not necessarily lead to statistical outcomes. Although there have been studies in the past \citep{Smith1976} on the role of exchange reactions in vibrationally inelastic collisions in systems like (Cl, HCl), we are not aware of any such studies on the role of exchange reaction in rotationally inelastic scattering processes.  
 \
 Since the two helium atoms involved in the collision
are indistinguishable, it follows that the actual observable would be a sum of the cross section values for the exchange and the inelastic processes. Therefore, we shall include the
H$^+$-exchange channel (${j^{\prime} > }$ 0, for ${v^{\prime}}$ = 0) to the inelastic channels  obtained from the PS-PES when discussing  the inelastic processes in comparison with the data from the RR-PES, which will not have the H$^+$-exchange option within its dynamical treatment of a purely rigid rotor  molecular partner. The possible differences between the final rates obtained with the two methods is indeed one of the interesting results from the present study.
\begin{figure}[!ht] 
	\begin{center} 
		\begin{tabular}{cc} \vspace*{-6mm} \resizebox{8.55cm}{!}{\includegraphics{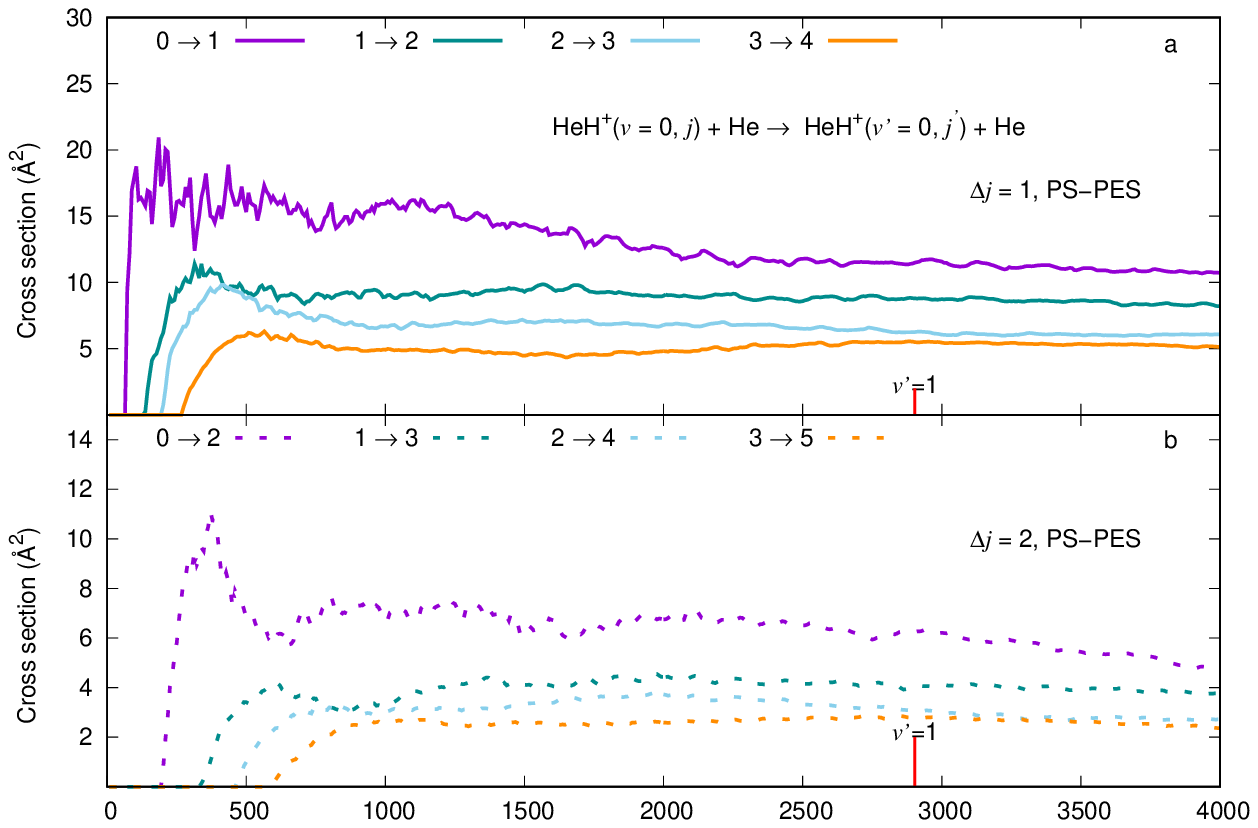}}\\ 
		\resizebox{8.9cm}{!}{\includegraphics[angle=270]{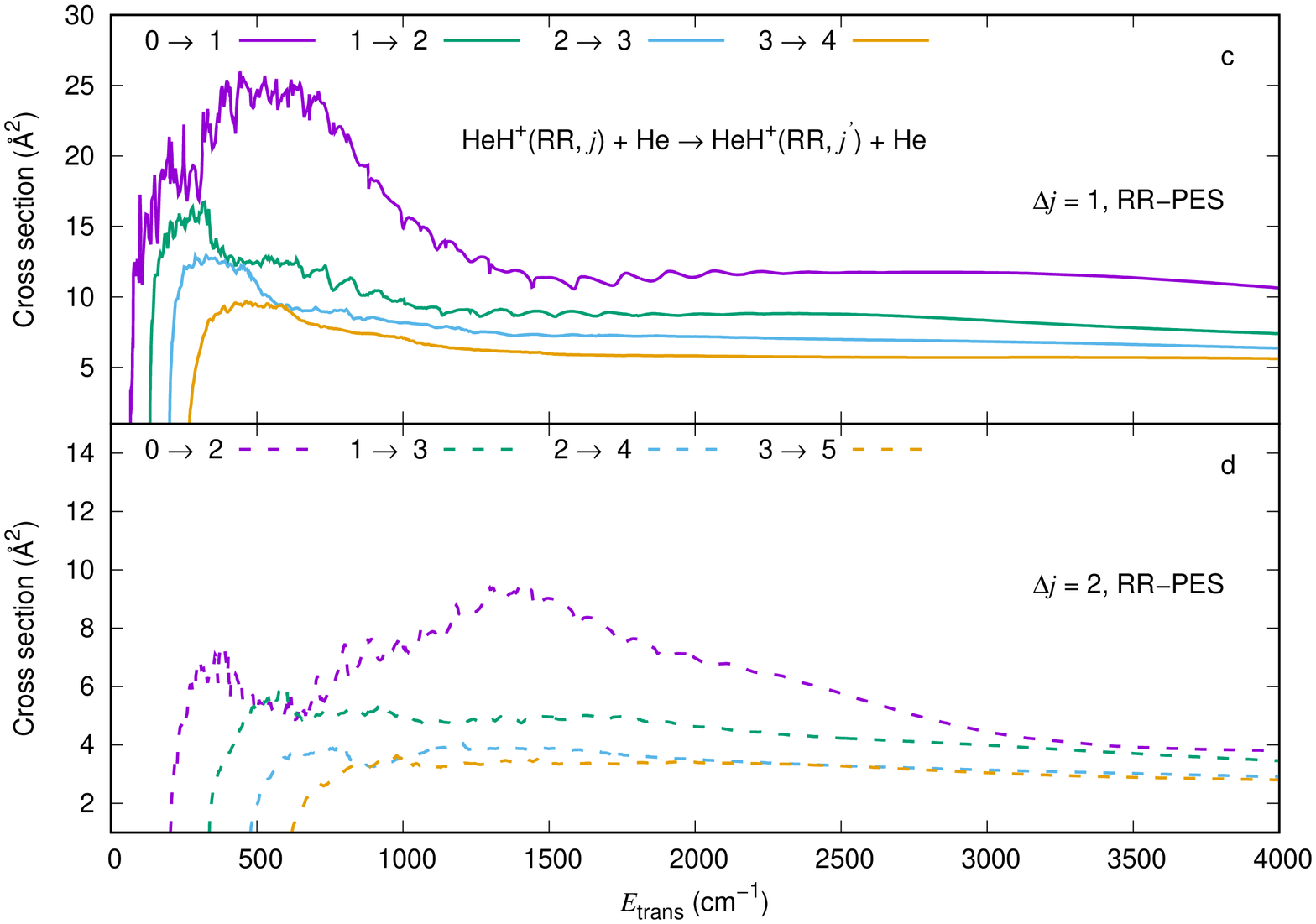}} \\ 
		\end{tabular} 
	\caption{Rotational excitation cross sections for $\Delta j$ = 1 and for $\Delta j$ = 2 transitions for HeH$^{+}$ ($j$ = 0-3) 
	collisions with He. The results in panels a) and b) are obtained via the PS-PES using the 3D interaction within the reactive code 
	ABC as discussed in the main text. The purely rotational inelastic results in panels c) and d) were obtained via the newly 
	constructed $ab$ $initio$ RR-PES using the present 2D interaction within the non-reactive code ASPIN also discussed in the main text. 
	The energy range of $E_{\textrm{trans}}$ values is the same in all four panels. The energy position for the openings of the first 
	excited vibrational level is pictorially marked in panels a) and b) to better clarify the relative energetics.} 
	\label{fig:fig5} 
	\end{center} 
\end{figure}
%
The state-to-state inelastic cross sections  involving different rotational states are  given in Figure \ref{fig:fig5},
where we show  the results for different $j$ (initial) and $j^{\prime}$ (final) states. 
The two upper panels in that Figure report the cross section results obtained using the ABC code and the PS-PES discussed in the previous Section, while the two lower panels report the same results when using the RR-PES of this work, also described in the previous Section, which follows the rigid rotor quantum dynamics.

We can see  that the energy threshold for the different
inelastic channels increases  with an increase in $j^{\prime}$, as is to be expected from the behaviour of the multipolar anisotropic coefficients discussed earlier. Furthermore, the magnitude of the cross sections decreases with an increase in the energy gap between $j$ and  $j^{\prime}$, as has been found for similar systems in many of our
earlier studies (see, for example, refs. \onlinecite{03MaBoGi, 08LoBoGi}).  The following additional comments can be made from a comparison of the two sets of calculations for the rotationally inelastic channels reported in Figure \ref{fig:fig5}:

$(i)$ the overall relative sizes of the different inelastic cross sections produced by the two different methods, which employ  different, but largely similar PESs as shown in the previous Section, turn out to be fairly close to each other in size and to also
exhibit very similar energy dependence over the observed range of energies. The inclusion of the H$^+$-exchange channels, therefore, is shown to make a fairly minor difference on the size of the final cross sections, apart from the energy regions just above thresholds;

$(ii)$  the purely inelastic cross sections (lower two panels) show a marked series of resonant structures in
the energy range up to about 500 cm$^{-1}$. Such resonant structures are also reproduced with similar
features by the calculations using the PS-PES interaction (upper two panels), where the inelastic flux also includes the H$^+$-exchange contribution as discussed earlier. Such similarities reflect the  similar strength of their anisotropic multipolar coefficients between the two PESs, as exhibited by the comparison given in
Figure \ref{fig:fig3}, while indicating the fairly minor role of the exchange channels with respect to the purely inelastic ones;

$(iii)$  the dominance of the anisotropy coupling induced by the $V_{\lambda}$ = 1 multipolar coefficient is also visible
when comparing the cross sections of the upper panels (panels a and c) with those reported by the ones below  (panels b and d):
the former are all larger than the latter over the whole energy range considered;

$(iv)$ as a general conclusion, we can say from the data in the panels of the above figure that both sets of calculations confirm
the  similarities between the anisotropy of the two PESs, and the fairly minor role played by the explicit inclusion of the H$^+$-exchange channels vs the rigid rotor dynamics. We further  see  that all inelastic cross sections turn out
to be rather large, indicating that the strong angular coupling terms in the potentials overcome the effects from having 
 rather large energy gaps between the involved rotational states, as shown in the previous Figure \ref{fig:fig2}.
 
To make sure that our conclusions are not dependent on the use of the two dimensional (2D) rigir rotor (RR) model or the choice of PES (RR-PES or PS-PES),
we have carried out 2D calculations using the PS-PES and the inelastic scattering code (ASPIN). The results are shown in Table~\ref{tab:table1}. Considering the fact that two different
PESs and two different dynamical models are used, the agreement between the results for different $\Delta j$ transitions can be considered excellent. The propensity for 
different $\Delta j$ transitions is also reflected very well in the two different models and the two different PESs used. 

\begin{table}[!ht]
\setlength{\tabcolsep}{8pt}
\renewcommand{\arraystretch}{1.2}
\caption{Comparison of representative results of inelastic cross section values (in $\AA^2$ units) for  HeH$^+$($j$ = 0) + He collisions, 
	obtained using the 2D (RR) model  and the two PESs at E$_{trans}$ = 1000 cm$^{-1}$. }
	\label{tab:table1}
\begin{tabular}{lccc}
\hline
 & 2D RR-PES & 2D(RR) PS-PES & 3D PS-PES \\
\hline
	$j'=1$ & 14.90 & 18.14 & 15.94 \\
	$j'=2$ &  7.47 &  7.42 &  6.90 \\
	$j'=3$ &  4.10 &  4.79 &  5.29 \\
\hline
\end{tabular}
\end{table}

 In conclusion, the present calculations unequivocally indicate that the present polar cation can efficiently exchange rotational energy by collisions with the He atoms present in the ISM environment  and therefore, as we shall further show below, can radiatively dissipate in that same environment the excess internal energy stored during those collisions.
 
 In order to extend the present comparison to the inelastic rate coefficients, and also  to compare our findings for the He partner  with what has been found with other partners that transfer energy by collisions to the cation rotational states,  we present further results in the following Section where the rate constants  are also presented.

\section{Comparing Helium-, Hydrogen- and electron-driven dynamics} \label{sec:resultd}

To assess the relative importance of  state-changing processes induced by 
 He, we compare their dynamical outcomes first  with  recently reported rate coefficients ($k$'s) 
for rotational energy transfer in HeH$^+$ by  collision with neutral
H , another important  component in the Interstellar Medium.
\begin{figure}[!ht]
        \begin{center}
                \begin{tabular}{cc}
                        \resizebox{9.2cm}{!}{\includegraphics[angle=270]{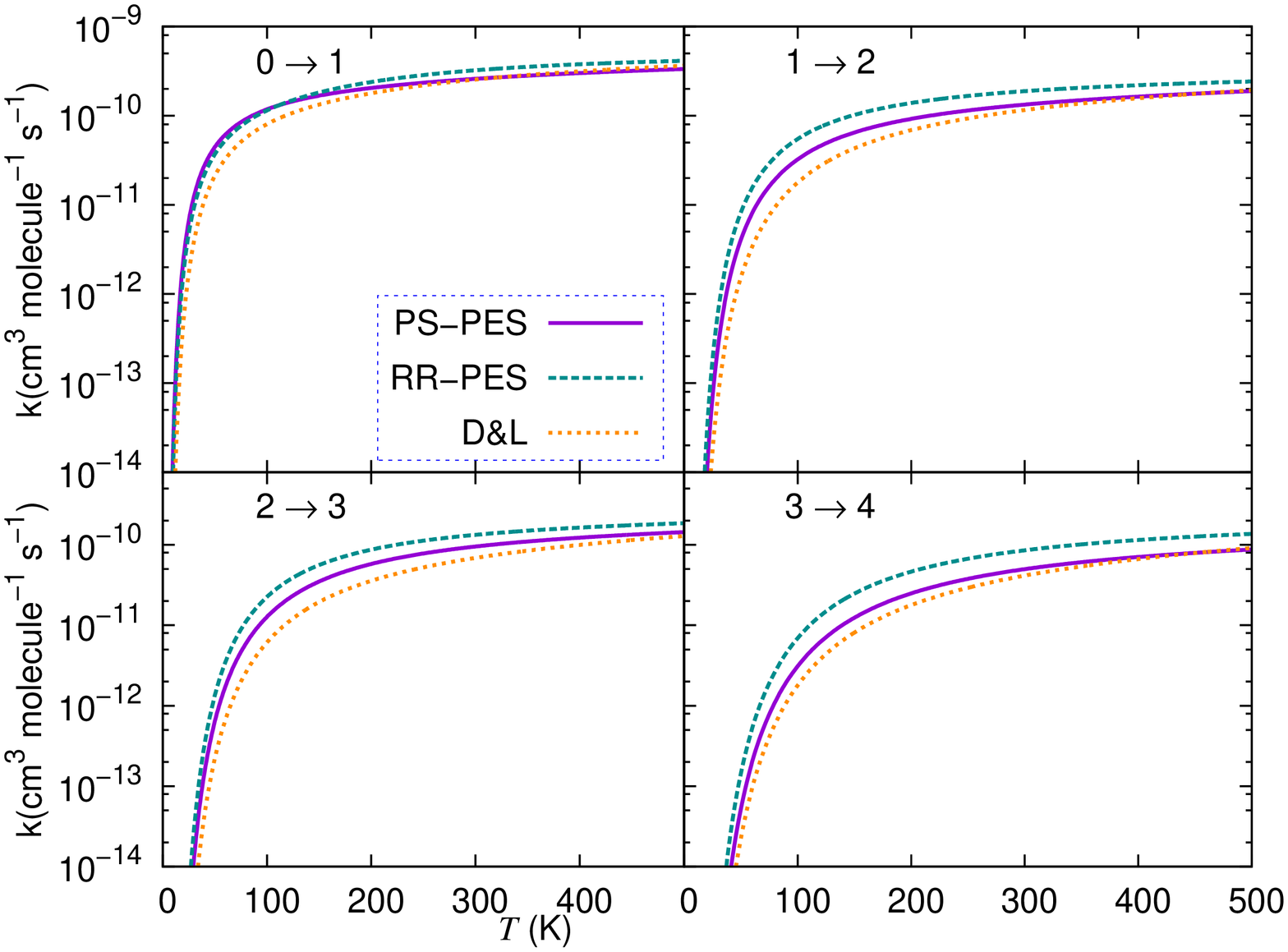}}\\
                        \resizebox{9.2cm}{!}{\includegraphics{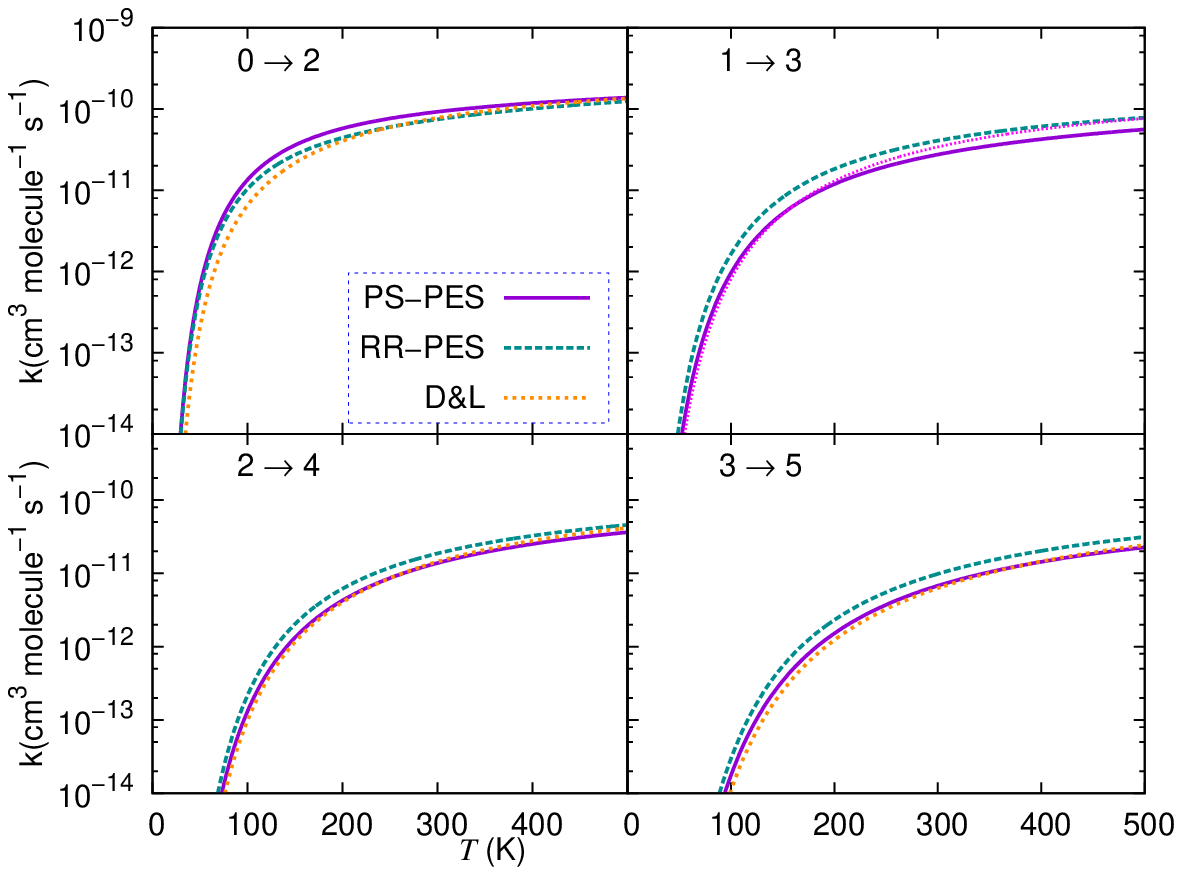}} \\
                \end{tabular}
                \caption{Temperature variation of the rate coefficients for rotational excitation of HeH$^+$ with $\Delta j$ = 1
                (upper four panels) and for $\Delta j$=2 (lower four panels)  obtained for HeH$^+$ collisions with He, using 
                the PS-PES and  the RR-PES of the present work. We include for comparison the earlier results from the work of \citet{Lique2020}
                on  HeH$^+$ ($v$ = 0, $j$ = 0, 1, 2 and 3) in collision with H atoms.}
                \label{fig:fig6} 
        \end{center}
\end{figure}
We have taken these rate coefficients as a function of temperature ($T$) from the earlier calculations
by \citet{Lique2020}, and have plotted our own computed rate coefficients for $\Delta j$ = 1 and 2 transitions for a direct comparison 
 in Figure \ref{fig:fig6}.  
\begin{figure}[!ht]
        \includegraphics[width=0.49\textwidth]{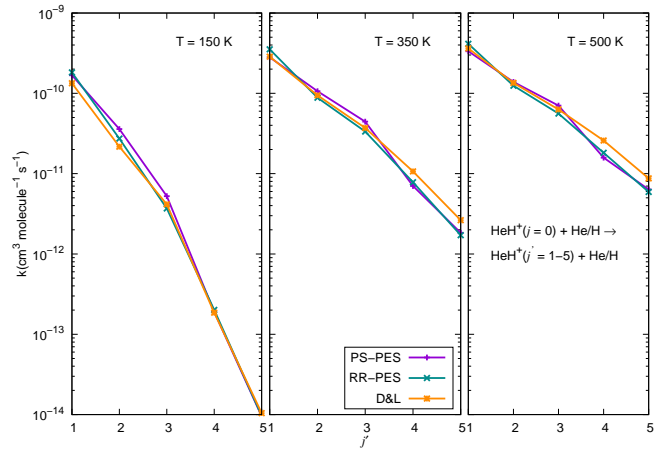}
        \caption{Rate coefficients for rotational excitation of HeH$^{+}$ ($v$ = 0, $j$ = 0) going to HeH$^{+}$($v^{'}$ = 0 and $j^{'}$
                $\neq$ 0) in collision with He. The present data are from  the quantum dynamics on the PS-PES and the  RR-PES discussed in the main text and are shown for three different temperatures.  Included for comparison are results from the earlier work of \citet{Lique2020} for HeH$^{+}$($v$ = 0, $j$ = 0) in  collision with neutral H.}
        \label{fig:fig7}
\end{figure}
It is clear from the four  upper panels of the Figure  \ref{fig:fig6},  which present the $\Delta j$ = 1 transitions,  
that the $k$ values for $j$ = 0 and 1 obtained using the rigid rotor dynamics on the RR-PES are 
fairly similar, and comparable in size and behaviour, to those we have obtained via  the 3D reactive dynamics using the PS-PES.  The  panels reporting additional transitions from higher initial $j$ states also confirm this similarity of rate coefficient values and of temperature dependence for the calculations via the two present PESs.   Interestingly, we also note that the purely rotationally inelastic results using the RR-PES  actually yield slightly larger rate
coefficients over the entire temperature ($T$) range than those obtained from the 3D reactive dynamics with  the PS-PES. This is in spite of the fact that the latter results  also include  the contributions from 
the  experimentally indistinguishable H$^+$-exchange route to the inelastic processes. 
\
Furthermore, the observables from either of the presently employed  PES are  clearly  very similar in  size to those obtained from collisions of the present cation with neutral H atoms, also reported in the same panels.

The  results  obtained for the state-changing processes with $\Delta j$ = 2 transitions, 
 presented in the four lower panels of the same figure, show that all state-to-state rate coefficients are
 uniformly smaller than those involving  $\Delta j$ = 1, as would be expected. Here again we see that all the inelastic rate coefficients for He turn out to be similar in size, albeit uniformly slightly larger than those
 where H is the collisional partner, thus underscoring the fact that one should consider both atomic projectiles  when modeling the kinetics of energy transfer paths involving   the present cation. In other words, the interesting new result from the present study is clearly the fact that He atoms should be considered at the same level of importance as the H  atoms when collisionally exciting rotational states of the title molecule.
 
 Another interesting way of looking at the relative efficiency of such inelastic rates is shown by the panels of Figure \ref{fig:fig7}. We see there that both our sets of results for the He partner  show  again very similar  excitation rate coefficients to those reported
 earlier by \citet{Lique2020} for the H partner. The data shown are  for the excitations from the $j$ = 0 state  going to $j'$ =1 through 5  over the range of $T$ (0-500 K): the three panels present three intermediate $T$-values as examples. The efficiency of the rotational excitation  by collisions with H are, in any case, of the same order of magnitude as those we have found here  for the He  collision partner.
 We also note   that the size of the rate coefficients  for excitations
  ending into increasingly  higher final $j'$ states  become dramatically smaller: the large energy gaps between states of  the ionic rotor   largely control that their size dramatically decreases as the gap increases. We  see  that the move to higher temperatures also causes the excitation rates to become dramatically larger for the same types of excitation. That the decline of the rate coefficient values with increasing  energy gap
is more marked at the lower temperatures turns out to be a   common feature for both atomic partners and to all calculations involving the present molecular system.
\begin{figure}[!ht]
        \includegraphics[width=0.49\textwidth]{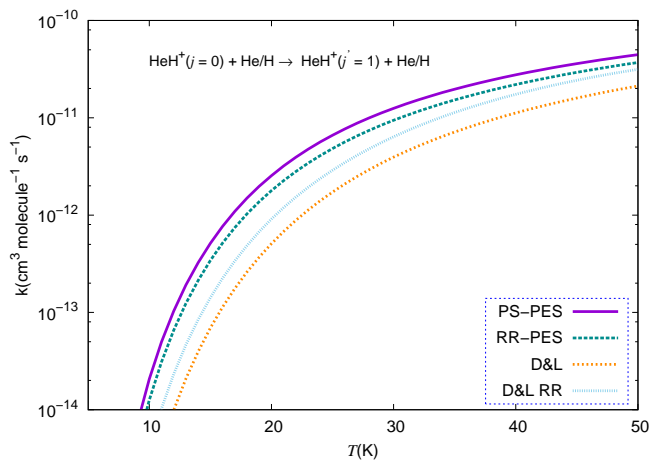}
        \caption{Temperature variation of the rate coefficient for the $j$ = 0 to $j'$ = 1 inelastic process on the PS-PES
                and on the RR-PES for HeH$^{+}$/He collisions at low temperatures. Included for comparison are results from the work
                of \citet{Lique2020} for HeH$^{+}$($v$ = 0, $j$ = 0) in collision with H. See main text for additional details.}
        \label{fig:fig8} 
\end{figure}

A further analysis is presented  for a range of  low temperatures up to 50 K in Figure \ref{fig:fig8}. We see that the  values of the rate constants obtained for the (HeH$^+$/ He) inelastic collisions,  using  both the RR-PES and the 3D reactive dynamics via the PS-PES, are  close to each other in size and behaviour over that range of temperatures, with the data from using the PS-PES being slightly larger than those obtained via the RR-PES. We know, however, (see Figure 9 ) that from around 100K and up to 500K the rate coefficients calculated via the RR-PES become larger than those produced via the PS-PES despite the former PES not including the H$^+$-exchange channel.
\
The rate coefficients obtained via the two PESs of the present work  show, in any case, fairly large values  for the rate coefficients: they are clearly larger than those reported by \citet{Lique2020},  for the collisions involving the H atom  and computed either with the same  rigid rotor dynamics of the present work  or  obtained from a dynamical treatment  where the H$_2$-formation channel is being  present in the quantum treatment: see \citet{Lique2020}.
  Therefore, we  can safely say that  the rigid-rotor type of rotationally inelastic dynamics produces somewhat larger inelastic rate coefficients in comparison to when  reactive channels are also considered in the  dynamics. In any event, the latter reactive flux is marginal in comparison with the size of the inelastic rate coefficients. Additionally, we find  that the efficiency of the rotational excitation of the  cation by He is larger than that obtained from collisions with the H atom in the same environmental conditions. \citep{Lique2020}  We also note that older results for the H atom as a collision partner,  \citet{KNSH19}  for the purely rotational excitation channels, were shown for comparison in an earlier study  by \citet{Lique2020} and  turned out to be markedly larger than those reported by \citet{Lique2020}. That   difference was there attributed to the lower quality of the PES employed in 
that earlier work. We have, therefore, omitted to report them  in the 
present analysis and suggest  that  the  rate coefficients  from the more  accurate PES should  preferably be taken into consideration for  comparative studies.

\begin{figure}[!ht]
        \includegraphics[width=0.345\textwidth,angle=270]{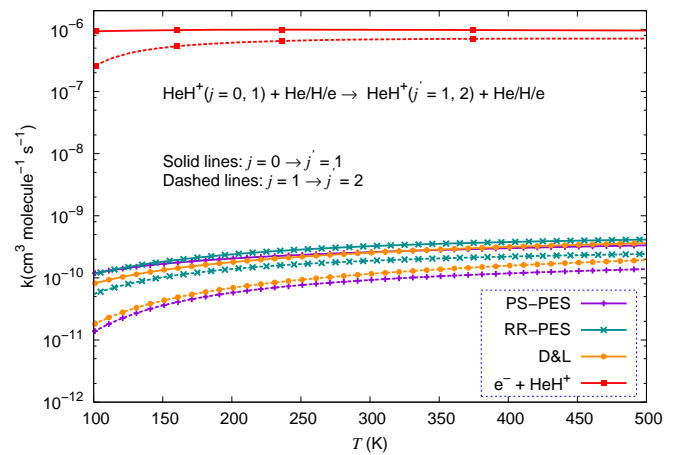}
        \caption{Comparison of rate coefficients for $\Delta j$ = 1 and $\Delta j$ =2 transitions  using either the PS-PES
        (lines with crosses) or the RR-PES (green curves with x signs) for HeH$^{+}$ ($j$ = 0) in collision with He. We also report
        the same type of rate coefficients of \citet{Lique2020} for HeH$^{+}$ in collision with H  and also those obtained by \citet{Hamilton2016} for the HeH$^{+}$ in collision with free electrons.}
        \label{fig:fig9}
\end{figure}
\begin{figure}[!ht]
        \includegraphics[width=0.49\textwidth]{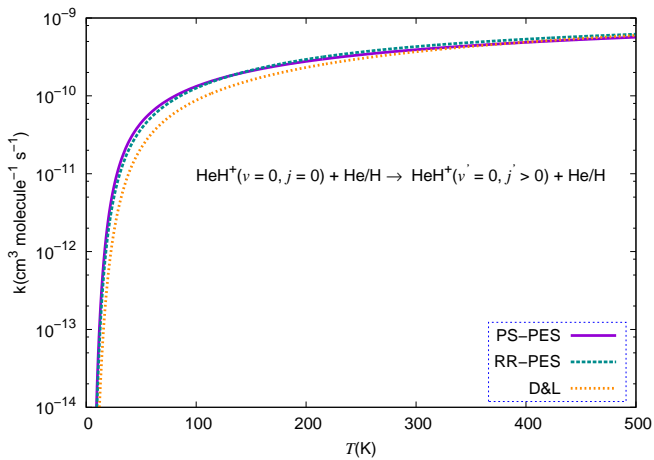}
        \caption{Comparison of the rate coefficients for rotationally inelastic (HeH$^{+}$($j$ = 0)/He) collisions summed over all
        state-to-state excitations into all $j´$ final states. The calculations were performed via the RR-PES and  the PS-PES discussed in the main text.
        Present results are compared with the same results reported earlier  by\citet{Lique2020} for HeH$^{+}$/H collisions.}
        \label{fig:fig10}
\end{figure}
The present results  dealing with
(HeH$^+$/ He) system are further compared with   those
reported by \citet{Hamilton2016} that involved HeH$^+$/$e^-$
collisions in the $T$ range of 100-500 K. The  comparison is shown in Figure  \ref{fig:fig9} together with the results from the neutral H  partner. We can see clearly that the  electron as a projectile is able to produce rotational excitation rate coefficients which are  about 3-4 orders of magnitude larger than those 
 caused  by collisions with either  He or H . Consistent with our earlier observations (vide supra), the rates for $\Delta j$ = 1
transition from $j$ = 0 are significantly larger than those starting  from $j$ = 1. The relative efficiency between individual collision processes naturally need to be correctly weighted with the different  densities estimated for these three different projectiles, so their actual role within kinetic evolutionary models requires the further evaluation of relative densities, as we shall briefly discuss  in the following Section for the case of He.

A further, perhaps more global, quantity can also  be employed  by 
showing the purely rotational inelastic rate coefficients where the individual state-to-state excitation processes from the $j$=0 initial level are summed over all the open excited levels as a function of temperature. The results are shown in Figure \ref{fig:fig10}. 
The present calculations for the He partner are given from the two different PESs employed here while the results for the H atom are marked as D\&L and are from ref. \onlinecite{Lique2020}. The very similar excitation efficiency shown by both partners is clearly evident in Fig. 10.

\section{Einstein coefficients and critical densities} \label{sec:erc}

Another important process for  the  decay of the internal rotational states of the ions in
the astrophysical  environments  is their interaction
with the surrounding radiation field.  The transition rates from an excited state $k$ can be written as a sum of stimulated and spontaneous 
emission rates as \citep{03BrCaxx}

\begin{equation}
\kappa^{em}_{k \to i} = \kappa^{sti}_{k \to i} + \kappa_{k \to i}^{spo} = A_{k \to i}(1 + \eta_{\gamma}(\nu,T))
\label{eq:tot_emi}
\end{equation}

Where $A_{k \to i}$ is the Einstein coefficient for spontaneous emission and 
$\eta_{\gamma}(\nu, T) = (e^{(h\nu/k_B T)} - 1)^{-1}$ is the Bose-Einstein photon occupation number.

The Einstein coefficient for dipole transitions is given as \citep{03BrCaxx}
\begin{equation}
A_{k \rightarrow i} = \frac{2}{3} \frac{\omega_{k \rightarrow i}^3 S_{k \rightarrow i}}{\epsilon_0 c^3 h (2 j_k + 1)}
\label{eq:dip_tran}
\end{equation}

Where $\omega_{i \to k} \approx 2B_0(j_i + 1)$ is the transition's angular frequency, and $S_{k \rightarrow i}$ is
the line strength. For pure rotational transitions, eq. \ref{eq:dip_tran} simplifies to 
\begin{equation}
A_{k \rightarrow i} = \frac{2}{3} \frac{\omega_{k \rightarrow i}^3}{\epsilon_0 c^3 h} \mu_0^2 \frac{j_k}{(2 j_k + 1)}
\label{eq:dip_tran_rot}
\end{equation}

Where $\mu_0$ is the permanent electric dipole moment of the molecule. 
In the present case the calculated value of 1.66 D has been employed. \citep{Dabrowski_Herzberg1977}
\begin{table}[!ht]
\setlength{\tabcolsep}{10pt}
\renewcommand{\arraystretch}{1.2}
\caption{Computed Einstein spontaneous emission coefficients $A_{j \to j'}$ for $^{4}$HeH$^{+}$ ($B_0$ =
 33.526cm$^{-1}$, $\mu = 1.66$ D), HD$^{+}$ ($B_{e}$ = 22.5 cm$^{-1}$, $\mu$ = 0.87 D)
and C$_{2}$H$^{-}$ ($B_{e}$ = 1.389 cm$^{-1}$, $\mu$ = 3.09 D).
All quantities in units of s$^{-1}$. The data for HD$^{+}$ and C$_{2}$H$^{-}$ are taken from ref. \onlinecite{20MaGiWe}.}.
\label{tab:EinA}
\begin{tabular}{cccc}
\hline
     Transition   & $^4$HeH$^+$ & HD$^+$  & C$_2$H$^-$   \\
\hline
$1 \to 0$ & 8.68$\times 10^{-2}$ & 7.2$\times 10^{-3}$ & 2.14$\times 10^{-5}$ \\
$2 \to 1$ & 8.33$\times 10^{-1}$ & 6.9$\times 10^{-2}$ & 2.05$\times 10^{-4}$ \\
$3 \to 2$ & 30.14$\times 10^{-1}$ & 2.5$\times 10^{-1}$ & 7.43$\times 10^{-4}$ \\
$4 \to 3$ & 74.09$\times 10^{-1}$ & 6.0$\times 10^{-1}$ & 1.83$\times 10^{-3}$ \\
$5 \to 4$ & 1.479$\times 10^{+1}$ & 1.2$\times 10^{0}$  & 3.65$\times 10^{-3}$ \\
\hline
\end{tabular}
\end{table}
A sample of the present results is collected in the  Table  \ref{tab:EinA}, where   other diatomic systems of interest in the ISM environment are also reported for comparison. Their properties are all taken from ref.  \onlinecite{20MaGiWe}. One should also note here that earlier calculations of the fuller range of rovibrational coefficents was presented by Engel et al. \citep{EDHT05}. Their results are close to ours, differing at most  by 10-15\%. Such difference is chiefly due to the differences in the potential curves employed by the  calculations and does not substantially change  our following discussion.
The present system, due to the large energy separations 
between rotational states, produces by far the largest values for the Einstein rate coefficients of 
spontaneous emission in the comparison shown in our Table  \ref{tab:EinA}. As we shall discuss further below, such large differences play an important 
role in the radiative dissipation of its internal energy in the low-density astrophysical environments of interest for our discussion.

The local thermodynamic equilibrium (LTE) assumption
holds whenever the population of excited levels is given by the  Boltzmann’s law. This
happens when the rate of spontaneous emission is significantly smaller than the rate of collisional
de-excitation and therefore can be neglected. This  means that
the density of gas should be significantly larger than some  critical value of that density so that the LTE assumption can be kept.
The above analysis is  made more specific by  the concept of  critical density  as recently mentioned, for instance,  in \citet{19LMSH}. The definition of a critical density is given as:

\begin{equation}
n_{\text{crit}}^i(T) = \frac{A_{ij}}{\sum_{j\neq i} k_{ij}(T) }
\label{eq:critD}
\end{equation}

\begin{figure}[!ht]
        \includegraphics[width=0.345\textwidth,angle=270]{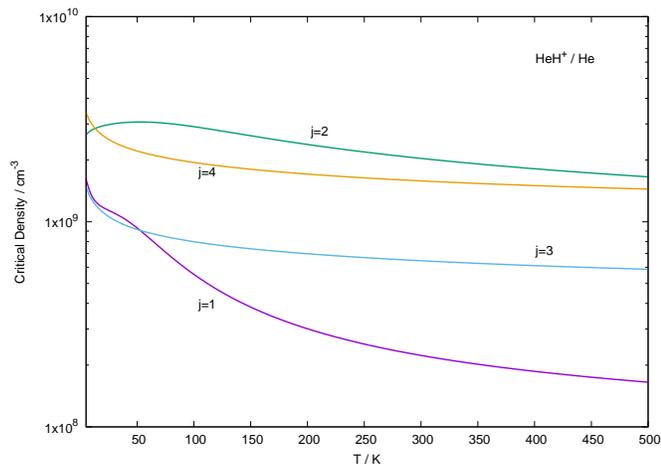}
        \caption{Computed critical densities for the HeH$^{+}$/He system, as defined in Equation eq. \ref{eq:critD}, over a range of temperatures from 5K and up to 500 K. See main text for further details.}
        \label{fig:fig11}
\end{figure}

Where the critical density  for any ith rotational level  is obtained by giving equal weight
to the effects of the collision-induced  and the spontaneous emission processes. We have taken the  rate coefficients discussed in the previous Sections, and using the RR-PES for both excitation and de-excitation processes, populating by collisions rotational levels up to $j$=9. We have also employed the computed spontaneous decay Einstein coefficients discussed here, samples of which are given by the above Table \ref{tab:EinA}. We report the results for the present evaluations in the panel of Figure  \ref{fig:fig11}, where the obtained critical density values are given along the Y-axis and the range of $T$ considered is along the X-axis. We clearly see there that the large values obtained for the critical densities in this case are mainly controlled by the very large spontaneous radiative coefficients discussed before.  

From  the variety of estimated baryon densities in the early universe environments discussed earlier, we have already mentioned
that the baryon density $n_b$  is proportional to the red shift value $z$  via the relationship: (1+$z$)$^3.$ \citep{Galli_Palla2013}  Hence, we can say that for values of  $z$ varying between 200 and 5000 the corresponding 
$n_b$ values have to vary between  
about 10$^{-1}$ cm$^{-3}$  up to about  10$^3$ cm$^{-3}$.  Therefore, we see  that the critical densities
associated to all the rotational levels we are considering  are  higher than the above 
estimated values for the baryon densities. This means that we should consider those populated states  
 not to be under the LTE conditions since the critical density values are all large enough to allow the molecules 
to radiate before they can collisionally de-excite. Under such conditions, therefore, to know accurately the
collision-driven rates calculated in our work would be important since the LTE conditions cannot be reliably employed.
More specifically, our present modeling of the critical densities deals chiefly with those associated to the He atoms. However, since the densities of the latter partner are  not very far from those for the H atom (about 90\% for H and about 10\% for He), our comparison between the two species, in relation to the more general baryon densities we are reporting, is still significant. It indicates that radiative decay of the rotationally excited states of the HeH$^+$ is going to be by far the dominant channel for internal energy release in the above environments.

To have shown that collisions can efficiently populate those excited rotational levels, which  in turn rapidly dissipate their stored energy by radiative emission, is therefore a significant result from the present calculations.

\section{Rotational Relaxation Kinetics in Ion Traps} \label{sec:rrk}

Since the data discussed in the previous Sections indicate that the collisional state-changing processes induced by He atoms among the rotational states of the HeH$^+$ cation are rather efficient processes, it is also interesting to further analyse the  role of  such new inelastic rates in  an entirely different physical situation, i.e. that associated with the trapping of the present cation under the much denser conditions of a cold ion trap where the He partner plays the role of the buffer gas. We have studied such processes many times in a variety of small molecular cations, so we shall not repeat the general discussion here but simply refer to those earlier publications \cite{17HeGiWe,18GoWeGi,18GiLaHe, 17KoGiHe}, while only a brief outline of the theory will be given below. We already know, in fact, that given the information we have obtained from the calculations , we are now in a position to try and follow the microscopic evolution of the cation's rotational state populations in a cold ion trap
environment by setting up the corresponding rate equations describing such 
evolution, induced by collisional energy transfers with the uploaded He gas, as described in various of our earlier studies \cite{19GoWeGi}:

\begin{equation}
	\label{eq:kin}
	\frac{d\mathbf{p}}{dt} = n_{He} \mathbf{k}(T)\cdot \mathbf{p}(t)
\end{equation}

Where the quantity $n_{He}$ indicates the density of the buffer gas loaded into  the trap, the He partner playing this time  the role of the collisional coolant . The vector $\mathbf{p}(t)$ contains the time-evolving 
fractional rotational populations of the ion partner's rotational state, p$_ j(t)$, from the initial distribution at t=t$_{initial}$, and the
matrix  $\mathbf{k}(T)$  contains the individual k$_{i\rightarrow j}(T)$ rate coefficients at the temperature of the trap's conditions. 
Both the p(t$_{initial}$) values and the collisional temperature T of the trap corresponding to the mean collisional energy between the partners
are quantities to be specifically selected in each computational run and will be discussed in detail in the modelling examples presented below. In
the present study we shall disregard for the moment the inclusion of the state-changing rates due to spontaneous radiative processes in the trap since, contrary to what happens under the conditions described in the previous Section, if we compare the critical densities shown by Figure \ref{fig:fig11} with those expected in traps,  we see them to be smaller than those of the collision-controlling,  more common buffer gas densities selected in the traps, as we shall show below. They are therefore not expected to have a significant effect under the more usual trap conditions \cite{17HeGiWe}.

We have chosen the initial rotational temperature of the trap's ions to be at 400 K, so that the vector's components at t=t$_{initial}$ are given
by a Boltzmann distribution at that chosen temperature. This was done  in order to follow the kinetics evolution over an extended range of time 
and also test the physical reliability of our computed state-changing collision rate constants. 

If the rate coefficients of the $\mathbf{k}(T)$ matrix satisfy the detailed balance between state-changing transitions, then as t$\rightarrow \infty$ 
the initial Boltzmann distribution will approach that of the effective equilibrium temperature of the uploaded buffer gas as felt by the ions in the trap.
 These asymptotic solutions correspond to the steady-state conditions  and can be obtained by solving the corresponding homogeneous form of eq. \ref{eq:kin} given
as: $d\mathbf{p} (t)/dt = 0$. We solved the homogeneous equations by using the singular-value  decomposition technique (SVD) \cite{17KoGiHe}, already 
employed by us in previous studies. 
The non-homogeneous equations \ref{eq:kin}, starting from our t$_{initial}$ 
of 400 K, were solved using the Runge-Kutta method for different  translational temperatures of the trap \cite{17KoGiHe,19GoWeGi}. 

Another useful indicator which could be extracted from the present calculations is the definition of a characteristic time, $\tau$, which can 
be defined as:

\begin{eqnarray}
	\label{tau}
	\left\langle E_{rot}\right\rangle (\tau)&&\, -\, \left\langle E_{rot} \right\rangle (t=\infty)\, =  \nonumber\\
	&& \frac{1}{e} \left(\left\langle E_{rot} \right\rangle(t=0) - \left\langle E_{rot} \right\rangle(t=\infty)\right)
\end{eqnarray}

The quantity $\left\langle E_{rot} \right\rangle$ represents the level-averaged rotational internal energy of the molecule in the trap after a 
characteristic time interval $\tau$ defined by equation \ref{tau}. It obviously depends on the physical collision frequency and therefore it depends on the $n_{He}$ value present in the trap \cite{17KoGiHe,19GoWeGi}. 

\begin{figure}[!ht]
        \includegraphics[width=0.49\textwidth]{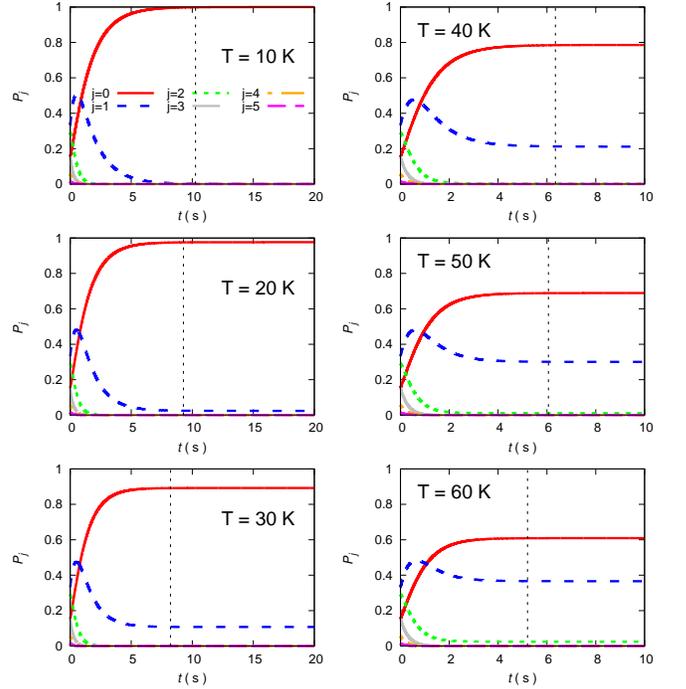}
        \caption{Computed time evolution  of the rotational state fractional populations in the cold trap. The buffer gas density was chosen to be of 10$^{10}$ cm$^{-3}$. The six panels show different choices for the trap temperatures and the vertical lines indicate when the steady-state population is reached in each situation. See main text for further details.}
        \label{fig:fig12}
\end{figure}

The results for the time evolution in a specific example of buffer gas density value (10$^{10}$ cm$^{-3}$) and for a temperature range between 10K and 60K are shown in the  six panels of Figure \ref{fig:fig12}. We present there the solutions of the kinetic eq.s  that were started at an initial T value of 400K, reporting in each panel the time lag needed to reach the steady-state populations at each temperature shown. The specific end time is located by the vertical line when the relative populations change no more by the 4th significant figure. Our data clearly indicate the effects of dealing with a light-atoms molecule where the energy gaps between levels are obviously rather large. At the lowest T values, in fact (e.g. for T at 10K and 20K) nearly all molecules reach the j=0 ground state within 10s, with only negligible populations of the excited states. As the temperature increases to 30K ,and then up to 60K, we see an increase of the number of molecules in the j=1 excited state, while however the most 
populated state remains the j=0 one, in contrast to what was seen in other cations where the ground state relative population did not dominate in the trap after the steady-state  \cite{18GoWeGi, 19GoWeGi} due to the smaller energy gaps existing for those systems between the lower rotational states.

The behaviour of the computed   $\tau$ values over a range of  trap densities possible in current cold trapped ion experiments, and over the range of temperatures also possible in those traps   \cite{17HeGiWe,18GoWeGi,18GiLaHe, 17KoGiHe}  is obviously linked to the form of eq.15, so that  we clearly  expect a scaling of the $\tau$ values with the increasing densities for the buffer gas:  four orders of magnitude of change for the $n_{He}$ will see a similar span of values for the characteristic times. They are also expected to change very little over the chosen range of T values.

\section{Present Conclusions} \label{sec:pconc}

 We have  obtained  a new potential energy surface from first principles, using quantum chemical methods with highly correlated functions as described in the earlier Section 2.  The molecular target was treated as a rigid rotor within the 2D description of the new interaction potential RR-PES and our findings were further compared with those from a  treatment that allowed the cation to vibrate and to undergo an H$^+$-exchange reaction using the earlier  PS-PES . Both potential functions were therefore employed to generate a wide variety of rotationally inelastic cross sections and to extract from them the corresponding inelastic rate coefficients pertaining to either purely rotational energy-transfer channels or the rotationally inelastic processes combined with the contributions from the H$^+$-exchange reactive channels. Our calculations found that the inelastic rates produced by the two different potential functions turned out to be very similar with each other and  clearly showed that the collisional excitation of the rotational internal states of HeH$^+$  interacting with He atoms is an important process, efficiently  yielding rotationally excited states of this cation under the ISM conditions modeled in this study. 
 
 We have further compared our present results with  earlier calculations in the literature which used other likely partners like neutral H atoms and free electrons  under the same external conditions we have employed for the He partner. We compare  a variety of different  collision efficiency indicators, clearly showing for the first time  that He and H are inducing rotational excitation processes with very similar efficiency, with He consistently turning out to be the more efficient partner. These findings therefore suggest that both neutral atoms have to be considered when including inelastic processes within the chemistry of early universe kinetic models. 
 
We have further calculated Einstein Coefficients fto estimate spontaneous decay of the cation´s excited rotational states  into a range of lower  levels. These coefficients  turn out to be among the largest found for simple molecular cations with light atoms and therefore clearly indicate that, given the expected baryonic densities at different redshift values from  current models (see  ref. \onlinecite{Galli_Palla2013}),  the critical densities required for the collisional paths to compete with the radiative paths are not likely to be present in the expected interstellar environments where this molecule has been detected. This indicates  that LTE conditions are not likely to be achieved for the molecular internal temperatures and therefore  the rapid radiative decay of rotational states down to the ground level will be the dominant cooling path for the present molecular ion. 
 
 To test an entirely different physical environment we have also run simulations of the time evolution of the rotational states of the present cation when confined in a cold trap where the He partner plays the role of the cooling buffer gas. The density values of the latter are obviously much larger than in the  ISM conditions, so that we can have quantitative information on the efficiency of this buffer gas as a coolant  under laboratory conditions and over a much more dense environment. The results  indicate the efficiency of the collisional cooling channels and the selective role on the relative populations played by the large energy gaps which exist  in this light-atom molecular cation.
 
 The present  work has thus provided from first principles a broad set of dynamical data and rotationally inelastic rate coefficients which  allow for a more realistic modeling of the chemical and baryonic evolution kinetics in the astrophysical environments and also for a quantitative evaluation of the  efficiency of the collisional cooling paths under cold ion trap laboratory conditions.

\section{Supplementary Material}

The multipolar coefficients for the Legendre expansion of the new RR-PES for HeH$^+-$He, the computed proton-exchange reaction cross sections, 
the computed inelastic rate coefficients from both the PESs employed in this work, and the convergence tests for the 3D reactive 
calculations (S1) are available as Supplementary Material to the present publication.

\section{Acknowledgements}

 FAG and RW  acknowledge the financial support of the Austrian FWF agency
through research grant n. P29558-N36. One of us (L.G-S) further thanks MINECO (Spain) for the awarding of grant PGC2018-09644-B-100. We are grateful to A. N. Panda for providing the potential energy subroutine for the PS-PES. We are also very grateful to F.Lique and B.Desrousseaux for generously providing  us with all  the numerical results published in their  paper  ready for comparison with our present calculations. 

\section{Data Availability}

The data that supports the findings of this study are available within the article and its supplementary material.

\nocite{*}
\bibliography{he2hp}



\end{document}


\title{Energy-transfer Quantum Dynamics of HeH$^+$  with He atoms: Rotationally Inelastic Cross Sections and Rate Coefficients}

\author{F. A. Gianturco}
\affiliation{Institut f\"{u}r Ionenphysik und Angewandte Physik, Universit\"{a}t Innsbruck\\
Technikerstr. 25
A-6020, Innsbruck, Austria}

\author{K. Giri}
\affiliation{Department of Computational Sciences, Central University of Punjab,\\
Bathinda 151001 India}

\author{L. Gonz\'{a}lez-S\'{a}nchez}
\email{lgonsan@usal.es}
\affiliation{Departamento de Qu\'{i}mica F\'{i}sica, University of Salamanca\\
 Plaza de los Ca\'{i}dos sn,
  37008, Salamanca, Spain}

\author{E. Yurtsever}
\affiliation{Department of Chemistry, Koç University \\
Rumelifeneriyolu, Sariyer
TR 34450, Istanbul, Turkey}

\author{N. Sathyamurthy}
\affiliation{Indian Institute of Science Education and Research Mohali\\
SAS Nagar, Manauli 140306 India}

\author{R. Wester}
\affiliation{Institut f\"{u}r Ionenphysik und Angewandte Physik, Universit\"{a}t Innsbruck\\
Technikerstr. 25\\
A-6020, Innsbruck, Austria}

\date{\today}

\begin{abstract}
\hspace*{6cm}{\textbf {Supplementary File}}
\end{abstract}

\maketitle
%

\renewcommand\thefigure{S\arabic{figure}}
\setcounter{figure}{0}
\begin{figure}[!ht]
        \begin{center}
\vspace*{-44mm} 
                \begin{tabular}{cc}
                       \vspace*{-34mm} 
                        \resizebox{19.5cm}{!}{\includegraphics[angle=270]{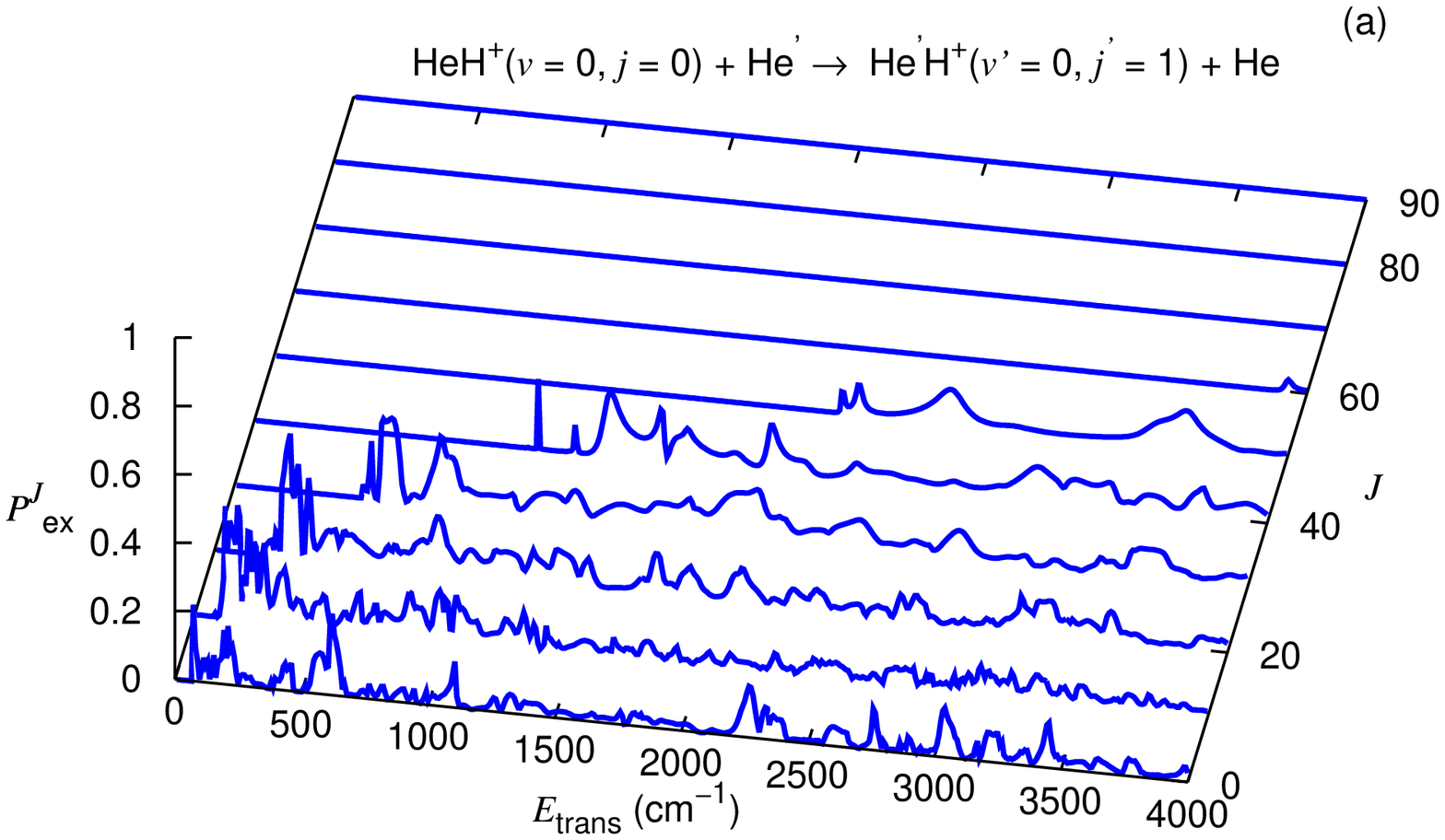}}\\
                        \resizebox{19.5cm}{!}{\includegraphics[angle=270]{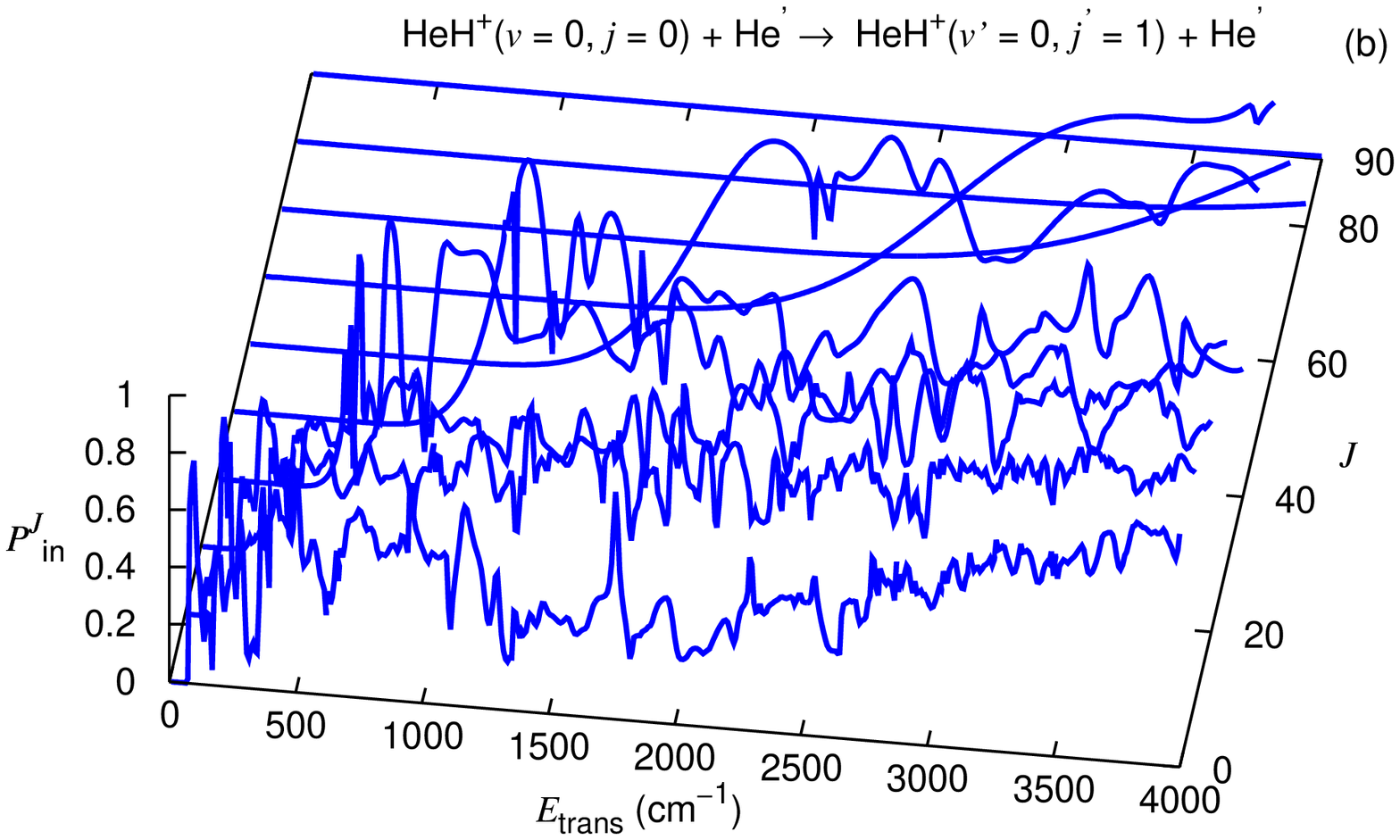}} \\
                \end{tabular}
                \caption{Illustration of the convergence of (a) exchange reaction probability and (b) inelastic transition probability 
for HeH$^{+}$ ( $v$ = 0, $j$ = 0 $\rightarrow$ ${v^{\prime}}$ = 0, ${j^{\prime}}$ =1) with respect to $J$ over a range of 
$E_{\textrm{trans}}$.}
                \label{fig:S1}
        \end{center}
\end{figure}